\documentclass[aps,prc,superscriptaddress,reprint]{revtex4-1}
\usepackage{graphicx} 
\usepackage{epstopdf}
\usepackage{dcolumn} 
\usepackage{bm}
\usepackage{hyperref}
\hypersetup{
    colorlinks=true,
    linkcolor=blue,
    filecolor=gray,
    urlcolor=blue,
    citecolor=blue,
}
\begin{document}
\title{To reach neutron-rich heavy and superheavy nuclei by multinucleon transfer reactions with radioactive isotopes}
\author{Peng-Hui Chen}
\affiliation{Institute of Modern Physics, Chinese Academy of Sciences, Lanzhou 730000, China}
\affiliation{School of Nuclear Science and Technology, University of Chinese Academy of Sciences, Beijing 100190, China}
\author{Fei Niu}
\affiliation{School of Physics and Optoelectronics, South China University of Technology, Guangzhou 510640, China}
\author{Wei Zuo}
\affiliation{Institute of Modern Physics, Chinese Academy of Sciences, Lanzhou 730000, China}
\affiliation{School of Nuclear Science and Technology, University of Chinese Academy of Sciences, Beijing 100190, China}
\author{Zhao-Qing Feng}
\email{Corresponding author: fengzhq@scut.edu.cn}
\affiliation{School of Physics and Optoelectronics, South China University of Technology, Guangzhou 510640, China}

\date{\today}
\begin{abstract}
  The dynamical mechanism of multinucleon transfer (MNT) reactions has been investigated within the dinuclear system (DNS) model, in which the sequential nucleon transfer is described by solving a set of microscopically derived master equations. Production cross sections, total kinetic energy spectra, angular distribution of formed fragments in the reactions of $^{124,132}$Sn+ $^{238}$U/$^{248}$Cm near Coulomb barrier energies are thoroughly analyzed. It is found that the total kinetic energies of primary fragments are dissipated from the relative motion energy and rotational energy of the two colliding nuclei. The fragments are formed in the forward angle domain. The energy dependence of the angular spectra is different between projectile-like and target-like fragments. Isospin equilibrium is governed under the potential energy surface. The production cross sections of neutron-rich isotopes are enhanced around the shell closure.

\begin{description}
\item[PACS number(s)]
25.70.Jj, 24.10.-i, 25.60.Pj
\end{description}
\end{abstract}

\maketitle

\section{Introduction}

The synthesis of superheavy nuclei (SHN) has obtained much progress up to element Z=118 (Oganesson) in cold fusion reactions \cite{Mu15} and by $^{48}$Ca-induced reactions \cite{Og15} in terrestrial laboratories. However, the observed isotopes are positioned away the island of stability because of the deficiency of neutrons. For eliminating the problem, the fusion-evaporation reactions induced by radioactive nuclides or the multinucleon transfer (MNT) reactions might be potentional way to approach the island. Meanwhile, it remains a topical subject in various laboratories and appropriate separation and detection techniques are in development \cite{Hei05}. Due to the broader excitation functions of MNT products, it has the advantage that a wide region isotopes can be populated in one experiment while complete fusion reactions are selective for only a few isotopes at a given beam energy and experimental setting. On the other hand, the properties of neutron-rich heavy isotopes are crucial in understanding the origin of heavy elements from iron to uranium in r-process of astrophysics. Traditionally, the neutron-rich isotopes are produced via the different mechanism by the fission of tansactinide nuclides, projectile fragmentation and complete fusion reactions for the light and medium mass region. However, extending to the heavy mass domain and even to island of superheavy stability, it is limited by the neutron abundance of projectile-target systems in the fusion-evaporation reactions. The MNT becomes a way to produce the neutron-rich heavy isotopes in nuclear chart \cite{bib:1,bib:17,bib:18}.

Following the motivation for producing heavy new isotopes and approaching the neutron-rich SHN, several models have been developed for describing the transfer reactions, i.e., the dynamical model based on multidimensional Langevin equations \cite{bib:3,bib:1}, the time-dependent Hartree-Fock (TDHF) approach \cite{bib:10,bib:11,Gu18,Ji18}, the GRAZING model \cite{bib:4,bib:6}, the improved quantum molecular dynamics (ImQMD) model \cite{Zha15}, and the dinuclear system (DNS) model\cite{Fe09,bib:8}, etc. Some interesting issus have been stressed, e.g., the production cross sections of new isotopes, total kinetic energy spectra of transfer fragments, structure effect on the fragment formation. There are still some open problems for the transfer reactions, i.e., the mechanism of preequilibrium cluster emission, the stiffness of nuclear surface during the nucleon transfer process, the mass limit of new isotopes with stable heavy target nuclides, etc. The extremely neutron-rich beams are favorable for creating neutron-rich heavy or superheavy nuclei owing to the isospin equilibrium \cite{Das94}. More discussions on the advantage of radioactive isotopes in MNT reactions may be referred in \cite{bib:9,mu15,lov19}.

The transfer reactions and deep inelastic scattering in damped heavy-ion collisions were investigated in experiments since 1970s, up to 15 transferred protons observed \cite{Hul77,Hil77,Das78,Gla79,Ess79,Fre79,Sch82,Moo86,Wel87,May85,Art71,Art73,Art74}. The fragment formation in the projectile-like and target-like region was investigated thoroughly. Recently, more measurements have been performed at different laboratories, e.g.,  the reactions of $^{136}$Xe+$^{208}$Pb \cite{Ko12,Ba15}, $^{136}$Xe+$^{198}$Pt \cite{Wa15}, $^{156,160}$Gd+$^{186}$W \cite{Ko17}, $^{238}$U+$^{232}$Th \cite{Wu18}. It has been shown that the MNT reactions are feasible for producing new isotopes. Sythesis of neutron-rich SHN beyond Z=105 via the MNT reactions have been arranged at the High Intensity Heavy-Ion Facility (HIAF) in the near future \cite{Ya13}.

In this work, the MNT reactions with the combinations of $^{124,132}$Sn+$^{238}$U/$^{248}$Cm are calculated with the DNS model. The article is organized as follows: In section II we give a brief description of the DNS model. Calculated results and discussions are presented in section III. Summary is concluded in section IV.

\section{Model description}

The DNS model has been used for the fusion-evaporation reactions, in which the competition of the quasifission and fusion to form heavy or superheavy compound nucleus is included \cite{Fe06,Fe07,Fe09b}. The DNS is assumed to be formed at the touching configuration of two colliding nuclei, in which the nucleon transfer is coupled to the dissipation of the relative motion energy and rotation energy. The cross sections of the primary fragments in the MNT reactions are given by
\begin{eqnarray}
\label{eq1}
\sigma _{pr} (Z_1, N_1, E_1) = && \sum ^{J_{max}} _{J=0} \sigma _{cap}(E_{c.m., J})    \nonumber \\
&& \int ^{B_{w-w}} _{B_{t-t}}f(B) P(Z_1, N_1, E_1, B) dB.
\end{eqnarray}
The capture cross section is evaluated by $\sigma _{cap} $= $\pi \hbar ^2 (2J +1)/ (2\mu E_{c.m.}) $$T(E_{c.m.}, J)$. The $T(E_{c.m.}, J)$ and $P(Z_1, N_1, E_1,B)$ are the transmission probability of the two colliding nuclei overcoming Coulomb barrier and fragment distribution probability, respectively. For two heavy nuclei, a parabola approximation of the interaction potential at the touching distance and the Hill-Wheeler formula \cite{Hw53} is used to evaluate the transmission probability. The barrier distribution approach is implemented into the model in order to include the contribution of all orientations and dynamical effect. The survived fragments are decay products of the primary fragments after emitting the particles and $\gamma$ rays in competition with fission \cite{Ch16}. The cross sections of the survived fragments are calculated by
\begin{eqnarray}
\sigma_{sur}(Z_1,N_1) = \sigma_{pr}(Z_1,N_1,E_1) \times W_{sur}(E_1, J_1, s).
\end{eqnarray}

The time evolution of the distribution probability P(Z$_1$,N$_1$,E$_1$,t) for a DNS fragment 1 with proton number Z$_1$, neutron number N$_1$ and with excitation energy E$_1$ is governed by the master equations as follows:
\begin{widetext}
\begin{eqnarray}
\label{eq2}
 \frac{d P(Z_1,N_1,E_1,t)}{d t}  = \sum _{Z'_1} W_{Z_1,N_1;Z'_1,N_1,}(t)  [d_{Z_1,N_1}P(Z'_1,N_1,E'_1,t) - d_{Z'_1,N_1}P(Z_1,N_1,E_1,t)]   + \nonumber \\ \sum _{N'_1} W_{Z_1,N_1;Z_1,N'_1,}(t)
[d_{Z_1,N_1}P(Z_1,N'_1,E'_1,t)  - d_{Z_1,N'_1}P(Z_1,N_1,E_1,t)].
\end{eqnarray}
\end{widetext}
Here W$_{Z_1,N_1;Z'_1 ,N_1}$ (W$_{Z_1,N_1;Z_1,N'_1}$ ) is the mean transition probability from the channel ($Z_1 ,N_1 ,E_1$) to ($Z'_1, N_1, E'_1$) [ or ($Z_1,N_1,E_1$) to ($Z_1,N'_1,E'_1$)], and $d_{Z_1,N_1}$ denotes the microscopic dimension corresponding to the macroscopic state ($Z_1,N_1,E_1$). Sequential one nucleon transfer is considered in the model and hence $Z'_1 =Z_1\pm 1$ and $N'_1 = N_1\pm1$. The transition probability is related to the local excitation energy and nucleon transfer, which is microscopically derived from the interaction potential in valence space as
\begin{eqnarray}
\label{eq3}
W_{Z_1,N_1;Z'_1,N_1,} = \frac{\tau_{mem}(Z_1,N_1,E_1; Z'_1,N_1, E'_1)}{d_{Z_1, N_1}d_{Z'_1, N_1} \hbar ^2 } \nonumber \\
 \times \sum _{ii'} |<Z'_1, N_1, E'_1, i' |V| Z_1, N_1, E_1, i>|^2
\end{eqnarray}
A similar formula for neutron transition coefficients. The relaxation time, typically being the order of 10$^{-22}$ s,  is evaluated by the deflection function method \cite{Ri79,Li81}.

The local excitation energy $E_1$ is determined by the dissipation energy from the relative motion and the potential energy surface of the DNS. The dissipation of the relative motion and angular momentum of the DNS is described by the Fokker-Planck equation \cite{Fe06,Fe07}. The energy dissipated into the DNS is expressed as
\begin{eqnarray}
\label{eq4}
E^{diss}(t) = E_{c.m.} - B - \frac{<J(t)>[<J(t)>+1]\hbar ^2}{2\zeta}     \nonumber
\\ - <E_{rad}(J,t)>
\end{eqnarray}
Here the $E_{c.m.}$ and B are the centre of mass energy and Coulomb barrier, respectively. The radial energy is evaluated from
\begin{equation}
\label{eq41}
 <E_{rad}(J,t)> = E_{rad}(J,0) \exp{(-t/ \tau _r)}.
\end{equation}
The relaxation time of the radial motion $\tau _r$= 5 $\times 10^{-22}$ s and the radial energy at the initial state $E_{rad}(J,0) = E_{c.m.} - B - J_i(J_i +1) \hbar ^2 / (2 \zeta _{rel})$. The dissipation of the relative angular momentum is described by
\begin{equation}
\label{eq5}
<J(t)> = J_{st} +(J_i - J_{st}) \exp(-t/ \tau _J).
\end{equation}
The angular momentum at the sticking limit $J_{st} = J_i \zeta _{rel}/ \zeta _{tot}$ and the relaxation time $\tau _J = 15 \times 10^{-22}$ s. The $\zeta _{rel}$ and $\zeta _{tot}$ are the relative and total moments of inertia of the DNS, respectively. The initial angular momentum is set to be $J_i = J$ in the following work. In the relaxation process of the relative motion, the DNS will be excited by the dissipation of the relative kinetic energy.

\begin{figure}
\includegraphics[width=1.\linewidth]{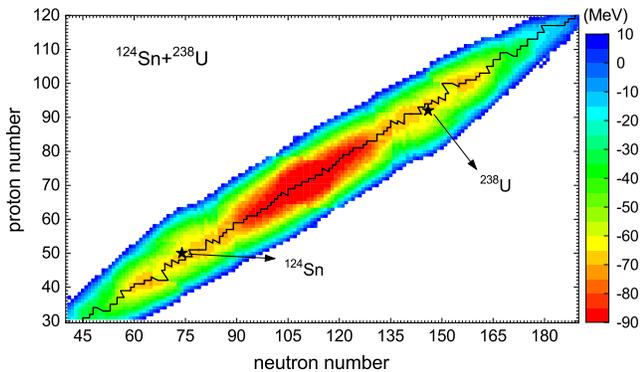}
\caption{\label{fig1} (Color online) Potential energy surface in the reaction of $^{124}$Sn+$^{238}$U and black line corresponding to the minimum value of each isotopic chain.}
\end{figure}

The local excitation energy is determined by the excitation energy of the composite system and the potential energy surface (PES) of the DNS. The PES is evaluated by
\begin{equation}
\label{eq6}
U(\{\alpha\})=B(Z_{1},N_{1})+B(Z_{2},N_{2})-B(Z,N)+V(\{\alpha\}),
\end{equation}
which satisfies the relation of $ Z_{1}+Z_{2}=Z $ and  $ N_{1}+N_{2}=N$ with the $Z$ and $N$ being the proton and neutron numbers of composite system, respectively. The symbol ${\alpha}$ denotes the quantities of $Z_{1}$, $N_{1}$, $Z_{2}$, $N_{2}$, $J$, $R$, $\beta_{1}$, $\beta_{2}$, $\theta_{1}$, $\theta_{2}$. The $B(Z_{i},N_{i}) (i=1,2)$ and $B(Z,N)$ are the negative binding energies of the fragment $(Z_{i},N_{i})$ and the composite system $(Z,N)$, respectively. The $\beta_{i}$ represent the quadrupole deformations of the two fragments and are taken the ground-state values. The $\theta_{i}$ denote the angles between the collision orientations and the symmetry axes of the deformed nuclei. Shown in Fig. \ref{fig1} is the PES in the tip-tip collisions of $^{124}$Sn+$^{238}$U. The DNS fragments towards the mass symmetric valley release the positive energy, which are available for nucleon transfer. The spectra exhibit a symmetric distribution for each isotopic chain. The valley in the PES is close to the $\beta$-stability line and enables the diffusion of the fragment probability.

\begin{figure}
\includegraphics[width=1.\linewidth]{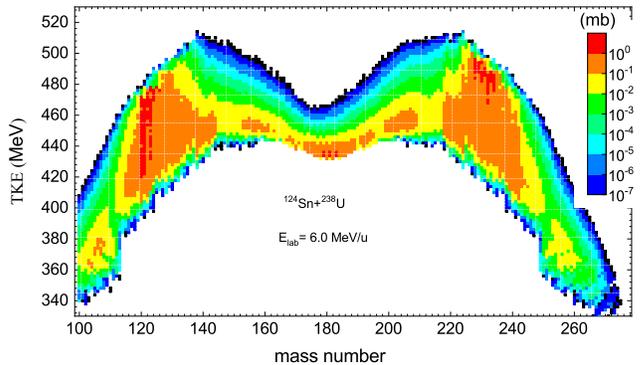}
\caption{\label{fig2} (Color online) Mass distribution of total kinetic energy of the primary binary fragments in the reaction of $^{124}$Sn+$^{238}$U at the incident energy E$_{lab}$ = 6 MeV/nucleon.}
\end{figure}

The total kinetic energy (TKE) of the primary fragment is evaluated by
\begin{equation}
\label{eq7}
TKE(A_{1}) = E_{c.m.} + Q_{gg}(A_{1}) - E^{diss}(A_{1}),
\end{equation}
where $Q_{gg} = M_P + M_T - M_{PLF} -M_{TLF}$ and $E_{c.m.}$ being the incident energy in the center of mass frame. The mass $M_P$, $M_T$, $M_{PLF}$ and $M_{TLF}$ correspond to projectile, target, projectile-like fragment and target-like fragment, respectively. The mass spectra of TKE is calculated as shown in Fig. \ref{fig2} in the reaction of $^{124}$Sn+$^{238}$U. More broad TKE dissipation is pronounced in the range of PLFs and TLFs. The formation of DNS fragments tend to the symmetric pathway (quasifission process).The spectra exhibit a symmetric mass distribution because of the structure in the PES.

The emission angle of the reaction products is helpful for arranging detectors in experiments. We use a deflection function method to evaluate the fragment angle which is related to the mass of fragment, angular momentum and incident energy. The deflection angle is composed of the Coulomb and nuclear interaction as \cite{Wo78,Ch18}
\begin{equation}
\label{eq8}
\Theta(l_i) = \Theta(l_i)_C+\Theta(l_i)_N
\end{equation}
The Coulomb scattering angle is given by the Rutherford function. The nuclear deflection angle is evaluated by
\begin{equation}
\label{eq9}
\Theta(l_i)_N = \beta \Theta(l_i)_C^{gr}  \frac{l_i}{l_{gr}} \left(\frac{\delta}{\beta} \right) ^{l_i / l_{gr}}.
\end{equation}
Here $\Theta_C^{gr}$  is the Coulomb scattering angle at the grazing angular momentum with $l_{gr} = 0.22 R_{int} [A_{red}(E_c.m. - V(R_{int}))]^{1/2}$. The $A_{red}$ and $V(R_{int})$ correspond to the reduced mass of DNS fragments and interaction potential at the distance $R_{int}$ of the entrance channel, respectively. The $\delta$ and $\beta$ are parameterized by fitting the deep inelastic scattering in massive collisions as
\begin{eqnarray}
\label{eq10}
\beta = &&  75 f(\eta) + 15,   \qquad  \eta <375      \nonumber       \\
&& 36 \exp(-2.17\times 10 ^{-3} \eta),   \qquad  \eta \geq 375
\end{eqnarray}

\begin{eqnarray}
\label{eq11}
\delta = && 0.07 f(\eta) + 0.11,     \qquad  \eta <375        \nonumber   \\
&&  0.117 \exp(-1.34\times 10 ^{-4} \eta),  \qquad  \eta \geq 375
\end{eqnarray}
and
\begin{equation}
\label{eq13}
f(\eta) = [1 + \exp{\frac{\eta-235}{32}}]^{-1}
\end{equation}
where $\eta = \frac{Z_1Z_2e^2}{\upsilon}$, and $\upsilon = \sqrt{\frac{2}{A_{red}}(E_c.m. - V(R_{int}))}$.

\section{Results and discussion}

\begin{figure}
\includegraphics[width=1.\linewidth]{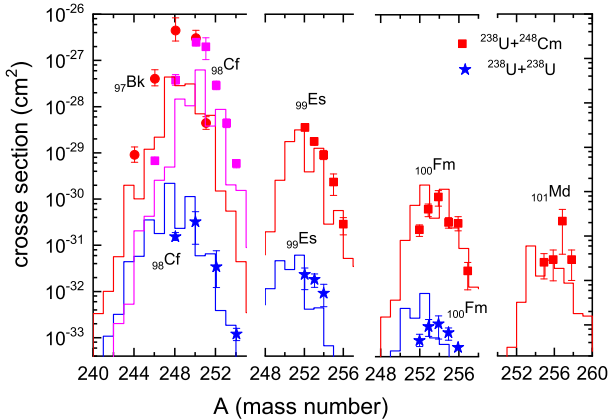}
\caption{\label{fig3}(Color online) Production cross sections of transcurium isotopes in the $^{238}$U+$^{238}$U reaction and $^{238}$U+$^{248}$Cm reaction at $E_{lab}=$ 7.0 MeV/u and compared with the available experimental data at GSI with error bars \cite{Kr13}.}
\end{figure}

The damped collisions of two actinide nuclei were investigated and motivated for producing superheavy nuclei in 1970s at Gesellschaft f\"{u}r Schwerionenforschung (GSI) \cite{Hul77,Hil77,Das78,Gla79}. Recently, the data were collected for investigating the MNT reactions, in particular for the transactinide production \cite{Kr13}. As a test of the DNS model, we calculated the isotopic cross sections in the reactions of $^{238}$U + $^{238}$U/$^{248}$Cm which are shown in Fig. \ref{fig3}. The isotopic yields are well reproduced with the model. It is obvious that the production cross section rapidly decreases with the actinide charge number. Up to 18 transferred nucleons were measured. The $^{248}$Cm based reactions are favorable for the transfermium isotope production owing to less nucleon transfer. To create SHN via MNT reactions, the heavy target nuclide are needed. The neutron shell closure is available for enhancing the transfer cross section in two actinide nuclide collisions. A bump structure of the isotopic yields around N=162 was predicted \cite{Fe09}, which is favorable for the neutron-rich SHN production.

The fragment yields in the MNT reactions are related to the emission angle in the laboratory system. It has been observed that the clusters formed in massive transfer reactions are emitted anisotropically \cite{Ji80}. Accurate prediction of the polar angle structure for the MNT fragments is helpful for managing the detector system in experiments. Shown in Fig. \ref{fig4} is the angular distribution of the primary MNT fragments produced in the reaction of $^{124}$Sn+$^{238}$U. The fragments are emitted in the forward region. The energy dependence of the projectile-like fragments (PLFs) in the mass region A=114-134 and target-like fragments (TLFs) with the mass number of A=228-248 is opposite. The PLFs are emitted towards the forward angle region when increasing the incident energy. The emission angles of the maximal yields for PLFs and TLFs are close at the energy of 6.5 MeV/nucleon.

\begin{figure}
\includegraphics[width=1.\linewidth]{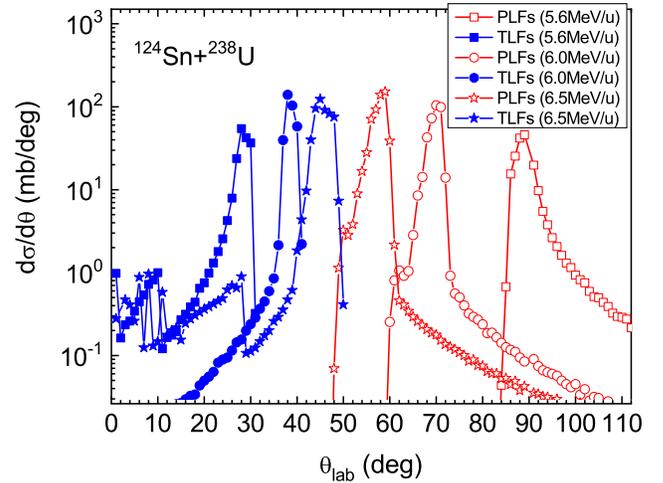}
\caption{\label{fig6}(Color online) Angular distributions of the Sn-like (blue lines) and the U-like products (red lines) in the laboratory frame at different energies.}
\end{figure}

Neutron-rich Sn isotopes can be generated by the asymmetric fission of actinide nuclide, for instance the new radioactive beam facility, Beijing Isotope Separation On Line (BISOL). The contour plot of primary and secondary fragments in collisions of $^{124,132}$Sn+$^{238}$U at the energy of 6 MeV/nucleon are calculated as shown in Fig. \ref{fig4}. The black zigzag line, straight line and pentagram symbols correspond to the minimal values of each isotopic chain in the PES, linking line of entrance system and position of projectile and target nuclides, respectively. The primary fragments are produced on the neutron-rich side. The de-excitation process moves the fragments to the $\beta$ stability line or even to the proton-rich domain. The nucleon transfer tends to the symmetric DNS fragments governed by the PES, in which the deformation, shell effect and odd-even phenomena influence the dissipation process. The diffusion of primary fragments reaches the transfermium isotopes and even close to superheavy element Ds (Z=110). Prompt de-excitation of primary fragments disenables the survival of SHN because of the low fission barriers. The PLFs and TLFs in the reaction $^{124}$Sn+$^{238}$U accumulate the neutron shell closure, i.e, around N=82 and 152. The entrance system in the reaction $^{132}$Sn+$^{238}$U is positioned on the valley of the PES, which enables the nucleon diffusion along the zigzag line. The $^{132}$Sn induced reactions are favorable to produce neutron-rich nuclei. Accurate estimations of the fission barrier for actinide and  transfermium nuclides are of importance for calculating the production cross section in the MNT reactions. Multidimensionally constrained covariant density functional approach was attempted to estimate the fission barrier of actinide nucleus \cite{Zh16}.

\begin{figure*}
\includegraphics[width=1\linewidth]{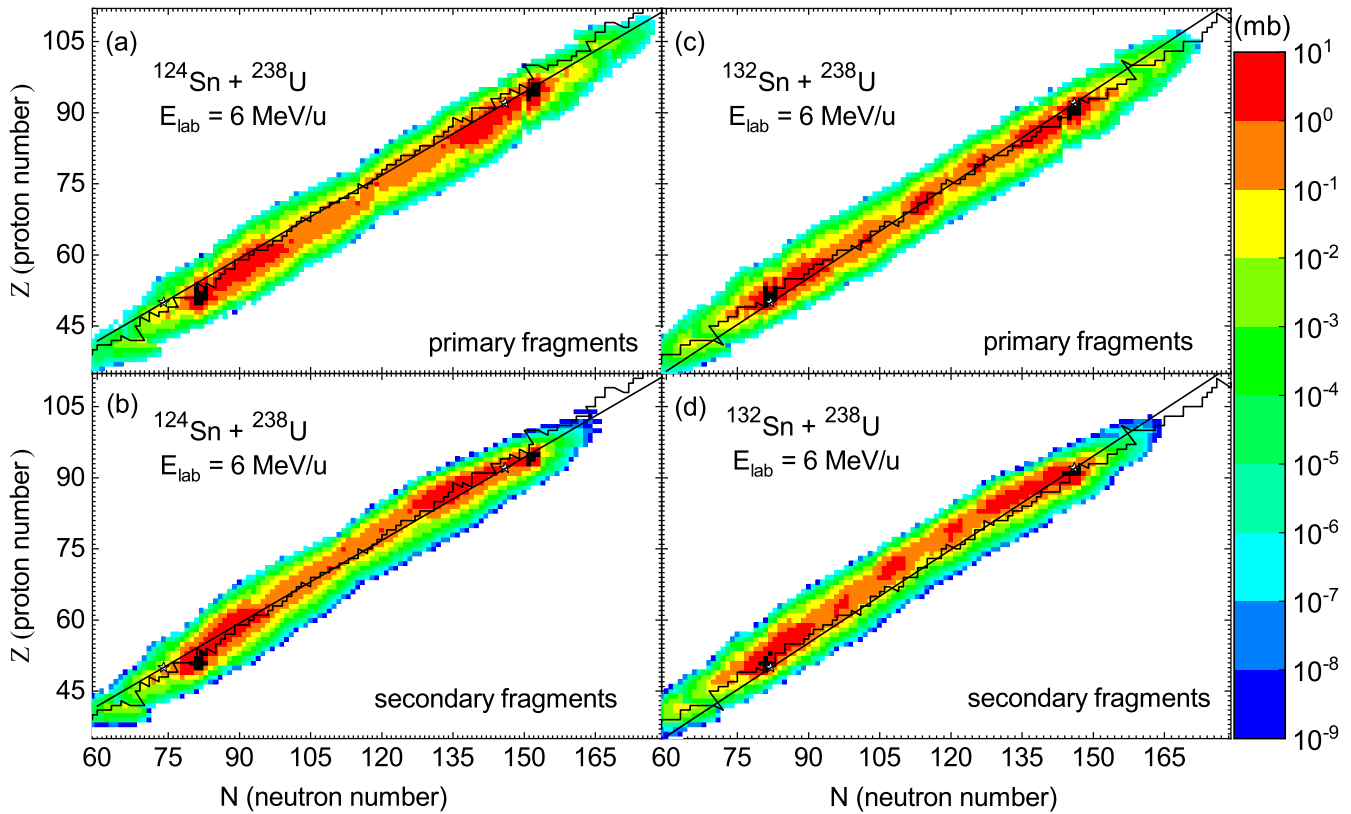}
\caption{\label{fig4}(Color online) Contour plot of production cross sections as functions of neutron and proton numbers of primary fragments and secondary fragments in collisions of $^{124}$Sn/$^{132}$Sn+$^{238}$U at the incident energy of 6 MeV/nucleon. The zigzag and straight lines corresponding to the minimal values of PES and to the neutron-proton line of target-projectile combination.}
\end{figure*}

The production cross sections of MNT fragments are related the reaction systems, in which the shell effect and isospin relaxation play significant roles on the fragment formation \cite{Fe17,Ni18}. The fragments in the preequilibrium process are produced around the projectile- or target-like region. More dissipations are available for creating the SHN and medium fragments. The nucleon transfer tends to the pathway along the valley in PES. Shown in Fig. \ref{fig6} are the isotopic spectra of production cross sections in the MNT reactions with $^{124}$Sn and $^{132}$Sn on $^{238}$U at the energy of 6 MeV/nucleon. The maximal yields move to the neutron-rich side for the proton pickup (left panels) and stripping (right panels) reactions with the bombarding nuclide $^{132}$Sn. The proton stripping reactions need to overcome the inner barrier in the PES and thus the cross sections drop rapidly with the proton number, e.g., $\mu b$ for 4 protons stripping from the bombarding nuclide. More nucleon transferring is needed for creating the transfermium isotopes.

\begin{figure}
\includegraphics[width=1\linewidth]{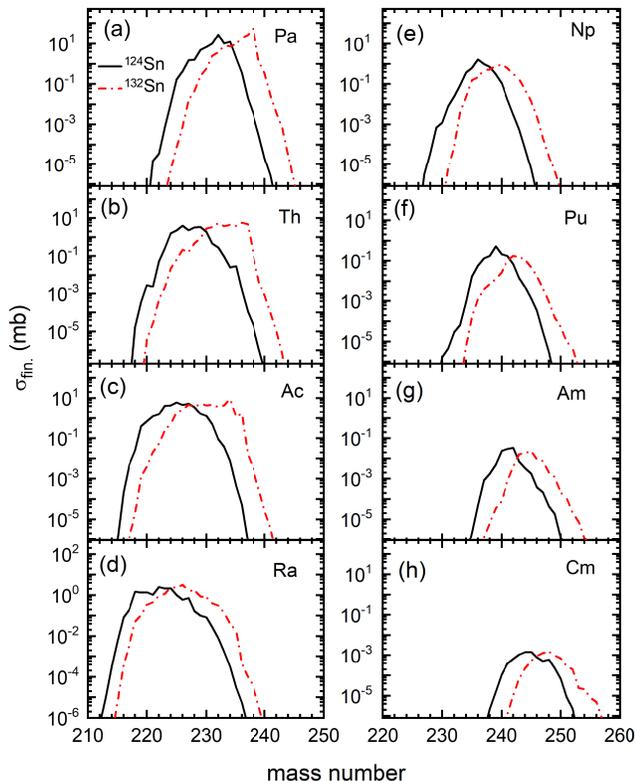}
\caption{\label{fig6}(Color online) Production cross sections of different channels in collisions of $^{124}$Sn and $^{132}$Sn on $^{238}$U at the incident energy 6 MeV/nucleon.}
\end{figure}

\begin{figure}
\includegraphics[width=1\linewidth]{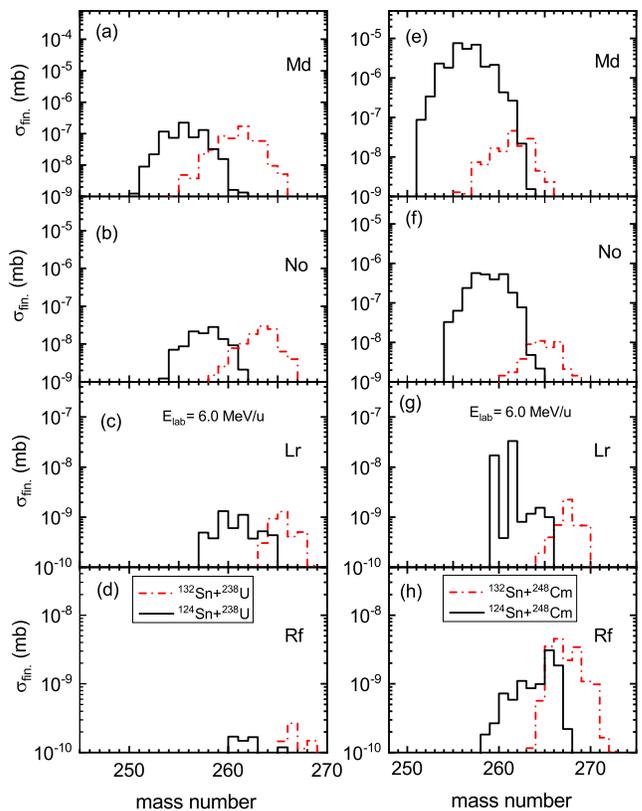}
\caption{\label{fig7}(Color online) Comparison of isotopic cross sections for producing elements of Z=101-104 with the $^{238}$U (left panels) and $^{248}$Cm based reactions (right panels) at the incident energy of 6 MeV/nucleon.}
\end{figure}

Figure \ref{fig7} is the cross sections for isotopes Z=101-104 in the MNT reactions of $^{124,132}$Sn + $^{238}$U/$^{248}$Cm. In the left panels it is shown that the cross sections induced by $^{124}$Sn has similar shapes and magnitudes with that induced by $^{132}$Sn, but the peak values shift towards the neutron-rich side for the latter case. However, for the spectra of $^{248}$Cm based reactions, the situation is somewhat complicate. The magnitudes are quite different for isotopes Z=101-103 with different projectiles. The peak values in $^{132}$Sn induced reactions are typically 1-2 order smaller than that induced by $^{124}$Sn. This is caused from that a shape transition from prolate to oblate ellipsoid takes place around A=110 for projectile-like isotopes Rh, Ru and Tc, which increase the interaction potential and enlarge the inner transfer barrier, and finally reduce the cross section significantly. The MNT reaction of $^{132}$Sn + $^{248}$Cm is promising pathway for producing the neutron-rich transfermium nuclides, in particular around the region of N=162. The nuclear spectroscopy and decay modes of the transfermium isotopes are the stepstone for investigating the SHN properties.

\begin{figure*}
\includegraphics[width=0.98\linewidth]{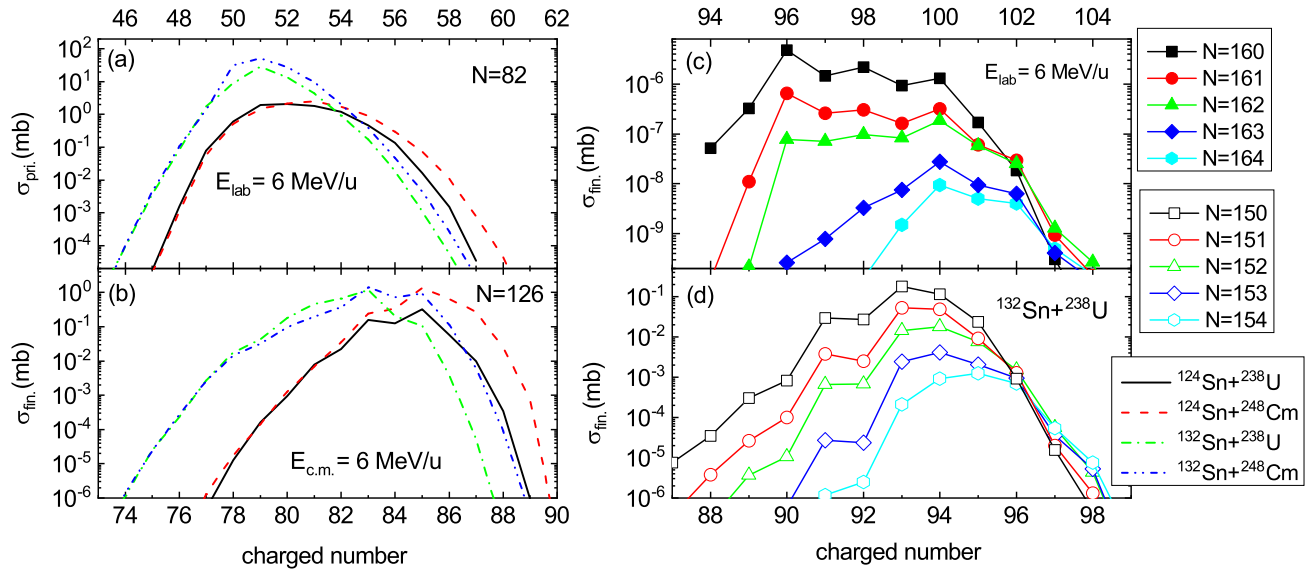}
\caption{\label{fig8}(Color online) Isotonic cross sections around neutron shell closure (N=82 and 126) in the reactions of $^{124,132}$Sn+ $^{238}$U/$^{248}$Cm (left panels) and the isotonic chains for the target-like fragments in the reaction of $^{132}$Sn+$^{238}$U (right panels).}
\end{figure*}

The isotonic cross sections in the four systems $^{124,132}$Sn + $^{238}$U/$^{248}$Cm at incident energy of 6 MeV/nucleon are calculated as shown in Fig. \ref{fig8}. It is obvious that the structure of the MNT yields is dependent on the projectile-target combinations. On the proton rich side, the production cross section is enhanced with enlarging the mass asymmetry of entrance system. The neutron-rich isotopes around N=82 and 126 are solely associated with the bombarding nuclide. The cross sections of neutron-rich nuclide are enhanced in the $^{132}$Sn induced reactions. The reaction systems reach the isospin equilibrium in the final evolution. The neutron to proton ratios are 1.548, 1.549, 1.603 and 1.605 for the reactions $^{124}$Sn+$^{248}$Cm, $^{124}$Sn+$^{238}$U, $^{132}$Sn+$^{248}$Cm and $^{132}$Sn+ $^{238}$Cm, respectively. The isotonic distribution around the subshell closures N=152 and 162 is shown in the right panels for the reaction $^{132}$Sn+$^{238}$U. The difference of isotonic cross sections is pronounced in the neutron-rich domain. It is caused from that the large N/Z ratios of isotopes in the neutron-rich region are away from the average value of the reaction system. However, the isospin ratios of heavy isotopes are close to the N/Z value of colliding system.

New isotopes might be produced via the MNT reactions. The cross sections are estimated as shown in Table \ref{tab1} for four systems of $^{124}$Sn+$^{238}$U, $^{124}$Sn+$^{248}$Cm, $^{132}$Sn+$^{238}$U and $^{132}$Sn+$^{248}$Cm at  incident energy E$_{lab}$= 6 MeV/u. Production cross section at the level of pb is feasible for measurements in laboratories. The number of new isotopes is indicated for the reaction systems. The new isotopic chains are broad with the $^{132}$Sn induced reactions. Further measurements are expected in the future experiments.

\begin{table*}
\caption{\label{tab1} Cross sections of unknown neutron-rich heavy and superheavy isotopes with proton number Z and mass number A, predicted by DNS model for the reactions of $^{124,132}$Sn + $^{238}$U/$^{248}$Cm around the Coulomb barrier energies from Nd to Rf. The last four columns show the number of new isotopes and their cross sections (in parentheses) which decrease with the increasing mass number.}
\begin{ruledtabular}
\begin{tabular}{ccccccc}

&Z (proton number) & A (mass number) & $^{124}$Sn+$^{238}$U(mb) & $^{124}$Sn+$^{248}$Cm(mb) &$^{132}$Sn+$^{238}$U(mb) & $^{132}$Sn+$^{248}$Cm(mb) \\
\hline
& Nd(Z=60)  & A$\geq$162    & 5 ($10^{-5} \sim10^{-9}$ )  & 5 ($10^{-4} \sim10^{-9}$ )  & 8 ($10^{-2} \sim10^{-9}$ ) & 8 ($10^{-3} \sim10^{-9}$ ) \\
& Pm(Z=61)  & A$\geq$164  & 6 ($10^{-4} \sim10^{-9}$ )  & 6 ($10^{-3} \sim10^{-9}$ )  & 9 ($10^{-1} \sim10^{-8}$ ) & 10 ($10^{-2} \sim10^{-9}$ ) \\
& Sm(Z=62)  & A$\geq$166  & 4 ($10^{-5} \sim10^{-7}$ )  & 4 ($10^{-5} \sim10^{-8}$ )  & 9 ($10^{-2} \sim10^{-9}$ ) & 9 ($10^{-2} \sim10^{-9}$ ) \\
& Eu(Z=63)  & A$\geq$168  & 4 ($10^{-5} \sim10^{-7}$ )  & 4 ($10^{-4} \sim10^{-8}$ )  & 10 ($10^{-2} \sim10^{-8}$ ) & 10 ($10^{-2} \sim10^{-8}$ ) \\
& Gd(Z=64)  & A$\geq$170  & 6 ($10^{-5} \sim10^{-7}$ )  & 5 ($10^{-4} \sim10^{-8}$ )  & 9 ($10^{-2} \sim10^{-8}$ ) & 9 ($10^{-2} \sim10^{-8}$ ) \\
& Tb(Z=65)  & A$\geq$172  & 8 ($10^{-4} \sim10^{-9}$ )  & 6 ($10^{-4} \sim10^{-9}$ )  & 11 ($10^{-1} \sim10^{-9}$ ) & 10 ($10^{-2} \sim10^{-9}$ ) \\
& Dy(Z=66)  & A$\geq$174  & 9 ($10^{-3} \sim10^{-9}$ )  & 8 ($10^{-3} \sim10^{-9}$ )  & 12 ($10^{-1} \sim10^{-9}$ ) & 11 ($10^{-1} \sim10^{-9}$ ) \\
& Ho(Z=67)  & A$\geq$176  &  8 ($10^{-3} \sim10^{-9}$ )  & 7 ($10^{-3} \sim10^{-9}$ )  & 11 ($10^{-1} \sim10^{-9}$ ) & 10 ($10^{-1} \sim10^{-9}$ ) \\
& Er(Z=68)  & A$\geq$178  & 7 ($10^{-3} \sim10^{-8}$ )  & 8 ($10^{-4} \sim10^{-9}$ )  & 11 ($10^{-1} \sim10^{-9}$ ) & 9 ($10^{-2} \sim10^{-9}$ ) \\
& Tm(Z=69)  & A$\geq$181  & 7 ($6\times10^{-3} \sim10^{-9}$)  & 7 ($10^{-3} \sim10^{-8}$ )  & 12 ($3\times10^{-1} \sim10^{-9}$ ) & 11 ($10^{-1} \sim10^{-9}$ ) \\
& Yb(Z=70)  & A$\geq$185  & 4 ($3\times10^{-5} \sim10^{-8}$ )  & 4 ($10^{-5} \sim10^{-9}$ )  & 10 ($10^{-3} \sim10^{-9}$ ) & 9 ($10^{-3} \sim10^{-9}$ ) \\
& Lu(Z=71)  & A$\geq$188  & 3 ($10^{-6} \sim10^{-9}$ )  & 6 ($10^{-6} \sim10^{-9}$ )  & 8 ($10^{-4} \sim10^{-9}$ ) & 8 ($10^{-4} \sim10^{-9}$ ) \\
& Hf(Z=72)  & A$\geq$190  &  1 ($\sim10^{-8}$ )  & 2 ($10^{-8} \sim10^{-9}$ )  & 7 ($10^{-4} \sim10^{-9}$ ) & 7 ($10^{-4} \sim10^{-9}$ ) \\
& Ta(Z=73)  & A$\geq$194  & 1 ($3.2\times10^{-8}$ )  & 2 ($10^{-8} \sim10^{-9}$ )  & 6 ($10^{-4} \sim10^{-9}$ ) & 7 ($10^{-4} \sim10^{-9}$ ) \\
&  W(Z=74)  & A$\geq$197  & 2 ($10^{-7} \sim10^{-9}$ )  & 3 ($10^{-7} \sim10^{-9}$ )  & 7 ($10^{-3} \sim10^{-9}$ ) & 7 ($10^{-4} \sim10^{-9}$ ) \\
& Re(Z=75)  & A$\geq$199  & 2 ($10^{-7} \sim10^{-9}$ )  & 2 ($10^{-7} \sim10^{-9}$ )  & 7 ($10^{-4} \sim10^{-9}$ ) & 8 ($10^{-4} \sim10^{-9}$ ) \\
& Os(Z=76)  & A$\geq$203  & 2 ($10^{-7} \sim10^{-9}$ )  & 2 ($10^{-7} \sim10^{-9}$ )  & 8 ($10^{-4} \sim10^{-9}$ ) & 8 ($10^{-4} \sim10^{-9}$ ) \\
& Ir(Z=77)  & A$\geq$205  &  $\textless 10^{-9}$  & 1 ($10^{-8}$ )  & 7 ($10^{-4} \sim10^{-9}$ ) & 7 ($10^{-4} \sim10^{-9}$ ) \\
& Pt(Z=78)  & A$\geq$208  & 2 ($10^{-7} \sim10^{-9}$ )  & 3 ($10^{-7} \sim10^{-9}$ )  & 9 ($10^{-3} \sim10^{-9}$ ) & 9 ($10^{-3} \sim10^{-9}$ ) \\
& Au(Z=79)  & A$\geq$211  & $\textless 10^{-9}$  & 1 ($ \sim10^{-9}$ )  & 7 ($10^{-4} \sim10^{-9}$ ) & 7 ($10^{-4} \sim10^{-8}$ ) \\
& Hg(Z=80)  & A$\geq$217  & 5 ($10^{-5} \sim10^{-9}$ )  & 5 ($10^{-4} \sim10^{-9}$ )  & 8 ($10^{-2} \sim10^{-9}$ ) & 8 ($10^{-3} \sim10^{-9}$ ) \\
& Tl(Z=81)  & A$\geq$218    & $\textless 10^{-9}$  & $\textless 10^{-9}$  & 4 ($10^{-6} \sim10^{-9}$ ) & 4 ($10^{-6} \sim10^{-9}$ ) \\
& Pb(Z=82)  & A$\geq$221  & $\textless 10^{-9}$  & $\textless 10^{-9}$  & 3 ($10^{-5} \sim10^{-9}$ ) & 4 ($10^{-5} \sim10^{-9}$ ) \\
& Bi(Z=83)  & A$\geq$225   & $\textless 10^{-9}$  & $\textless 10^{-9}$  & 3 ($10^{-5} \sim10^{-7}$ ) & 4 ($10^{-5} \sim10^{-9}$ ) \\
& Po(Z=84)  & A$\geq$227  & $\textless 10^{-9}$  & $\textless 10^{-9}$  & 3 ($10^{-6} \sim10^{-9}$ ) & 3 ($10^{-6} \sim10^{-9}$ ) \\
& At(Z=85)   &A$\geq$230   & $\textless 10^{-9}$  & $\textless 10^{-9}$  & 7 ($10^{-4} \sim10^{-9}$ ) & 5 ($10^{-5} \sim10^{-9}$ ) \\
& Rn(Z=86)   &A$\geq$232 & $\textless 10^{-9}$  & $\textless 10^{-9}$  & 5 ($10^{-5} \sim10^{-9}$ ) & 3 ($10^{-6} \sim10^{-9}$ ) \\
& Fr(Z=87)   &A$\geq$234  & 2 ($10^{-7} \sim10^{-9}$ )  & 2 ($ \sim10^{-8}$ )  & 5 ($10^{-4} \sim10^{-7}$ ) & 4 ($10^{-6} \sim10^{-9}$ ) \\
& Ra(Z=88)   &A$\geq$235  & 4 ($10^{-5} \sim10^{-9}$ )  & 3 ($10^{-6} \sim10^{-9}$ )  & 7 ($10^{-2} \sim10^{-9}$ ) & 5 ($10^{-4} \sim10^{-9}$ ) \\
& Ac(Z=89)   &A$\geq$238  & 5 ($10^{-4} \sim10^{-9}$ )  & 6 ($10^{-3} \sim10^{-9}$ )  & 9 ($1.6 \sim10^{-9} $ ) & 9 ($10^{-2} \sim10^{-9}$ ) \\
& Th(Z=90)   &A$\geq$240  & 2 ($10^{-5}, 10^{-7}$ )  & 4 ($10^{-4} \sim10^{-7}$ )  & 6 ($10^{-3} \sim10^{-9}$ ) & 7 ($10^{-3} \sim10^{-9}$ ) \\
& Pa(Z=91)   &A$\geq$242  & 3 ($10^{-5} \sim10^{-7}$ )  & 5 ($10^{-3} \sim10^{-7}$ )  & 7 ($10^{-1} \sim10^{-8}$ ) & 9 ($10^{-1} \sim10^{-8}$ ) \\
&  U(Z=92)   &A$\geq$244  & 4 ($10^{-5} \sim10^{-9}$ )  & 7 ($10^{-2} \sim10^{-8}$ )  & 8 ($10^{-1} \sim10^{-9}$ ) & 11 ($10^{-1} \sim10^{-9}$ ) \\
& Np(Z=93)   &A$\geq$245  & 3 ($10^{-6} \sim10^{-9}$ )  & 3 ($10^{-4} \sim10^{-9}$ )  & 7 ($10^{-2} \sim10^{-9}$ ) & 8 ($1 \sim10^{-9}$ ) \\
& Pu(Z=94)   &A$\geq$248  & 3 ($10^{-6} \sim10^{-9}$ )  & 2 ($10^{-6} \sim10^{-8}$ )  & 7 ($10^{-3} \sim10^{-8}$ ) & 7 ($10^{-3} \sim10^{-9}$ ) \\
& Am(Z=95)   &A$\geq$250  & 5 ($10^{-5} \sim10^{-9}$ )  & 4 ($10^{-3} \sim10^{-7}$ )  & 9 ($10^{-3} \sim10^{-8}$ ) & 9 ($26 \sim10^{-8}$ ) \\
& Cm(Z=96)   &A$\geq$253  & 3 ($10^{-6} \sim10^{-9}$ )  & 2 ($10^{-7} \sim10^{-9}$ )  & 7 ($10^{-4} \sim10^{-8}$ ) & 6 ($10^{-3} \sim10^{-8}$ ) \\
& Bk(Z=97)   &A$\geq$255  & 2 ($10^{-7} \sim10^{-9}$ )  & 3 ($10^{-6} \sim10^{-9}$ )  & 6 ($10^{-5} \sim10^{-7}$ ) & 7 ($10^{-4} \sim10^{-8}$ ) \\
& Cf(Z=98)   &A$\geq$257  & 1 ($1.8 \times10^{-9}$ )  & 3 ($10^{-7} \sim10^{-9}$ )  & 5 ($10^{-6} \sim10^{-9}$ ) & 6 ($10^{-5} \sim10^{-9}$ ) \\
& Es(Z=99)   &A$\geq$259  & $\textless 10^{-9}$  & 4($10^{-7} \sim10^{-9}$ )  & 6 ($10^{-7} \sim10^{-9}$ ) & 6 ($10^{-6} \sim10^{-9}$ ) \\
& Fm(Z=100)  &A$\geq$261  & $\textless 10^{-9}$  & 2($10^{-8} \sim10^{-9}$ )  & 4 ($10^{-7} \sim10^{-9}$ ) & 6 ($10^{-7} \sim10^{-9}$ ) \\
& Md(Z=101)  &A$\geq$261  & 1 ($1.3 \times10^{-9}$ )  & 3 ($10^{-7} \sim10^{-9}$ )  & 5 ($10^{-7} \sim10^{-9}$ ) & 5 ($10^{-8} \sim10^{-9}$ ) \\
& No(Z=102)   &A$\geq$265  & 1 ($2 \times10^{-9}$ ) )  & 4 ($10^{-7} \sim10^{-9}$ )  & 6 ($10^{-8} \sim10^{-9}$ ) & 8 ($10^{-8} \sim10^{-9}$ ) \\
& Lr(Z=103)   &A$\geq$267  & 2 ($ \sim10^{-9}$ )  & 5 ($10^{-8} \sim10^{-9}$ )  & 1 ($ \sim10^{-9}$ ) & 1 ($  \sim10^{-9}$ ) \\
& Rf(Z=104)   &A$\geq$269  & $\textless 10^{-9}$ & 2 ( $\sim10^{-9}$ )  & 0 ($\textless 10^{-9}$) & 5 ( $\sim10^{-9}$ ) \\
\end{tabular}
\end{ruledtabular}
\end{table*}

\section{Conclusions}

In summary, the production of neutron-rich isotopes via the MNT reactions has been investigated within the DNS model for the reaction systems of $^{124,132}$Sn+$^{238}$U/$^{248}$Cm around Coulomb barrier energies. The nucleon transfer takes place at the touching configuration of two fragments under the PES. The valley shape of the PES influences the formation of primary fragments and leads to the production of neutron-rich isotopes. The de-excitation process shifts the neutron excess of fragments towards the $\beta$-stability line. The isospin relaxation in the nucleon transfer is coupled to the dissipation of relative energy and angular momentum of colliding system. The available experimental data for two actinide nuclide collisions are well reproduced. The fragment yields are enhanced around the shell closure. The neutron-rich nucleus $^{132}$Sn (n/p=1.62) induced reactions are favorable to produce heavy neutron-rich isotopes around TLFs. The anisotropy emission of MNT fragments is associated with the incident energy of colliding system. The angular distribution of the PLFs is shifted to the forward region with increasing the beam energy. However, that of TLFs exhibits an opposite trend. The production cross sections of isotonic chains around neutron shell closure N=82 and 126 depend on the projectile-target combinations, in particular in the proton-rich domain. The difference between the isotonic cross sections around N=152 and 162 is pronounced in the neutron-rich region. The isotopic cross sections of Nd, Gd and Pb are related to entrance channel effect. Predicted numerous unknown neutron-rich nuclei from Z=60 to Z=104 predicted with cross section by DNS model within four reaction systems, which list in Table. The $^{132}$Sn induced reactions are available for the neutron-rich isotope production. Possible measurements are expected in future experiments.

\section{Acknowledgements}

This work was supported by the National Natural Science Foundation of China (11435014, 11675226, 11722546, 11975282) and the Talent Program of South China University of Technology.

\end{document}